\documentclass[12pt]{article}
\usepackage{epsfig}

\begin{document}

\begin{titlepage}
\begin{flushright}
astro-ph/0307463 \\ UFIFT-HEP-03-16 \\ CRETE-03-13
\end{flushright}

\vspace{1.5cm}

\begin{center}
\bf{IMPROVED ESTIMATES OF COSMOLOGICAL PERTURBATIONS}
\end{center}

\vspace{0.5cm}

\begin{center}
N. C. Tsamis$^{\dagger}$
\end{center}
\begin{center}
\it{Department of Physics, University of Crete \\
GR-710 03 Heraklion, HELLAS.}
\end{center}

\vspace{0.3cm}

\begin{center}
R. P. Woodard$^{\ddagger}$
\end{center}
\begin{center}
\it{Department of Physics, University of Florida \\
Gainesville, FL 32611, UNITED STATES.}
\end{center}

\vspace{0.7cm}

\begin{center}
ABSTRACT
\end{center}
\hspace*{.3cm}
We recently derived exact solutions for the scalar, vector,
and tensor mode functions of a single, minimally coupled
scalar plus gravity in an arbitrary homogeneous and isotropic
background. These solutions are applied to obtain improved
estimates for the primordial scalar and tensor power spectra
of anisotropies in the cosmic microwave background.

\vspace{1cm}

\begin{flushleft}
\end{flushleft}

\vspace{0.5cm}

\begin{flushleft}
$^{\dagger}$ e-mail: tsamis@physics.uoc.gr \\
$^{\ddagger}$ e-mail: woodard@phys.ufl.edu
\end{flushleft}

\end{titlepage}

\section{Introduction}

Mukhanov and Chibisov \cite{mukhanov} were the first to suggest
that quantum fluctuations during inflation produced the tiny
inhomogeneities needed to form the various cosmic structures
we observe currently -- as the result of gravitational collapse
over the course of more than 10 billion years. Early work on
the subject was also done by Hawking \cite{hawking}, by Guth
and Pi \cite{guth}, and by Starobinski\u{\i} \cite{starobinsky1}.
The formalism has since been described at length in a number
of review articles \cite{MFB,LL,LLK}. It has received much
attention recently owing to the unprecedented precision with
which the imprint of these fluctuations on the cosmic microwave
background radiation has been imaged by the WMAP satellite
\cite{WMAP1,WMAP2}. 

Much of the fascinating structure revealed by these measurements 
derives from processes which occurred long after the end of inflation, 
and are not the subject of this paper. Instead, we re-compute the 
primordial fluctuation spectrum which is the starting point for the 
analysis of subsequent processes. The justification is that we now 
have at our disposal the exact scalar and graviton mode functions 
upon which the calculation is based \cite{nctrpw2,nctrpw3}.
There has never been any doubt regarding the spacetime
dependence of the mode functions during the epoch of matter
domination in which the cosmic microwave background radiation
anisotropies accumulate. What was previously unavailable is
an exact expression for the normalization factor which the
mode functions build up during inflation.

Previous computations have been based on approximation schemes 
that were developed over the course of two 
decades. A key step in this effort was the introduction, by Stewart 
and Lyth, of the slow-roll Bessel function approximation \cite{SL}. 
However, Wang, Mukhanov and Steinhardt \cite{WMS} demonstrated that 
carrying this approximation to higher orders does not generally 
improve accuracy, while Martin and Schwarz \cite{MS2} showed that 
the technique's accuracy is not sufficient for comparison with 
precision experiments such as WMAP and PLANCK. Recent improvements 
\cite{SG,STG,HHJM,MS} have overcome these obstacles, at least for 
slow-roll inflation \cite{LLMS}, so the additional precision available 
from our exact solutions is probably not necessary for comparison 
with foreseeable data. But it is nice to have, and it is simple enough 
to construct exotic models in which the slow-roll paradigm breaks 
down completely. We shall study one in an appendix.

To fix notation, note that cosmologically relevant spacetimes are 
characterized by scale factor $a$:
\begin{equation}
ds^2 \; = \;
- \, dt^2 \, + \, a^2 (t) \, d{\vec x} \cdot d{\vec x}
\;\; . \label{ds2}
\end{equation}
Although not directly an observable, the ratio of its current
value $a_0$ to its value at past time $t$ is the cosmological
redshift experienced by light emitted at that time and received
now:
\begin{equation}
z(t) \; \equiv \;
\frac{a_0}{a(t)} \, - \, 1
\;\; . \label{z}
\end{equation}
Its logarithmic derivative defines the Hubble parameter
$H$ which measures the rate at which distant matter is
receding due to the expansion of the universe:
\begin{equation}
H(t) \; \equiv \;
\frac{{\dot a}(t)}{a(t)}
\;\; . \label{H}
\end{equation}
Its second time derivative enters into the deceleration
parameter $q$:
\begin{equation}
q(t) \; \equiv \;
- \, \frac{a(t) \; {\ddot a}(t)}{{\dot a}^2 (t)}
\; = \;
-1 \, - \, \frac{{\dot H}(t)}{H^2 (t)}
\;\; . \label{q}
\end{equation}
The weak energy condition implies that $q(t) \geq -1$;
inflation is characterized by $q(t) < 0$.

Quantum fluctuations are not especially big during inflation,
but they are enormously larger than afterwards. Therefore, we
can analyze the process using linearized quantum field theory.
Furthermore, the high degree of homogeneity and isotropy of
the inflationary geometry implies both that a fluctuation
can be characterized by its constant, co-moving wave vector
${\vec k}$ :
\begin{equation}
{\vec k} \; = \;
\frac{2\pi {\vec n}}{\lambda}
\;\; , \label{k}
\end{equation}
and that each fluctuation evolves independently. The
physical wave length $\lambda_{\rm ph}$ of a fluctuation
grows as the universe expands:
\begin{equation}
\lambda_{\rm ph} (t) \; = \; a(t) \; \lambda
\;\; , \label{lambdaph}
\end{equation}
where $\lambda$ is the co-moving wave length. During
inflation the Hubble radius $r_H$:
\begin{equation}
r_H (t) \; \equiv \;
H^{-1} (t)
\;\; , \label{rH}
\end{equation}
is approximately constant, whereas it grows more rapidly
than the scale factor after the end of inflation
{\it (see Figure 1)}. This variation of $r_H$ gives rise
to the two horizon crossings which characterize the
fluctuations of interest to us. They happen when:
\begin{equation}
Horizon \; Crossing
\quad \Longrightarrow \quad
\lambda_{\rm ph} (t) \; \equiv \;
r_H (t)
\;\; . \label{horcros1}
\end{equation}
First horizon crossing occurs during inflation. Before this
time the linearized fields oscillate with falling amplitude;
afterwards they are approximately constant. Second horizon
crossing occurs long after the end of inflation, indeed
after the emission of the cosmic microwave radiation.
Before second horizon crossing the fields are approximately
constant whereas they oscillate with falling amplitude
afterwards.

\begin{figure}
\centerline{\epsfig{file=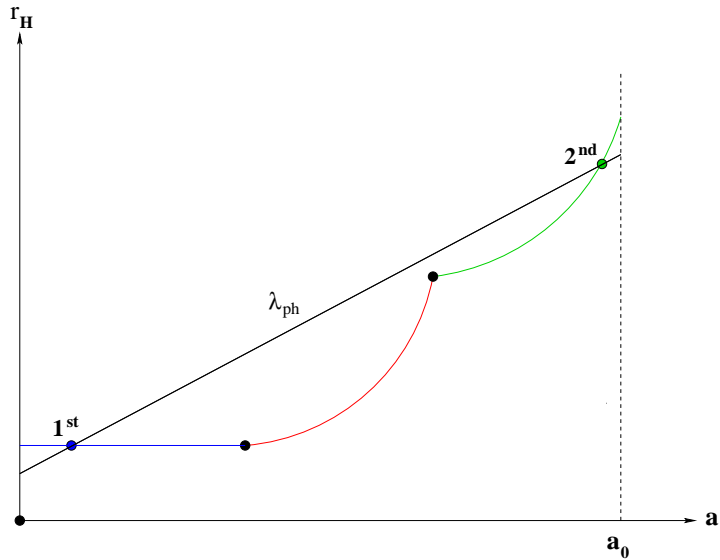,height=2.9in}}
\caption{\footnotesize The first and second horizon crossings
of the physical wavelength $\lambda_{\rm ph}$. The
\break \mbox{} \hspace{1.8cm}
Hubble radius $r_H$ is constant during inflation {\it (blue)},
behaves like $r_H \, \sim \, a^2$ during radiation {\it (red)},
and like $r_H \, \sim \, a^{\frac32}$ during matter domination
{\it (green)}. The present is at $a_0$. The graph is not
properly scaled.}
\end{figure}

We can be more precise by defining the dimensionless variable
$x$ which represents the physical wave number in Hubble units:
\begin{equation}
x(t, k) \; \equiv \;
\frac{k}{a(t) \, H(t)}
\;\; , \label{xdef}
\end{equation}
and in terms of which horizon crossing means:
\begin{equation}
Horizon \; Crossing
\quad \Longrightarrow \quad
x(t, k) \; = \; 1
\;\; . \label{horcros2}
\end{equation}
When the deceleration parameter $q$ is constant, which is
a good approximation during the dominant phases in the
history of the universe, the following relation is valid:
\begin{equation}
q(t) \; = \; {\bar q}
\quad \Longrightarrow \quad
x(t, k) \; = \;
x(t_i, k) \, \left( \,
\frac{1 + z(t_i)}{1 + z(t)} \, \right)^{\bar q}
\;\; . \label{qconstant}
\end{equation}
If we take the initial time instant $t_i$ to signify the
onset of inflation, it becomes apparent that during inflation
$x$ decreases with time: ${\bar q}$ is negative in this era
while $z$ is an ever decreasing function of time. Thus, the
modes of interest start with $x$ larger than 1 and achieve
condition (\ref{horcros2}) -- first horizon crossing -- as
the universe still inflates. After first horizon crossing,
the variable $x$ for these modes further decreases and becomes
very much less than 1. However, post-inflationary evolution
is characterized by positive ${\bar q}$ and, hence, $x$
increases during this era. The modes of physical interest
are such that condition (\ref{horcros2}) is satisfied again
-- second horizon crossing -- before the present.

Moreover, we note that significant fluctuations do not occur
for all fields, only for those which are both much lighter
than the inflationary Hubble parameter and also not conformally
invariant. These two requirements mean we need consider only
gravitons and light, minimally coupled scalars.

It is unnecessary to discuss invariant characterizations of
cosmological perturbations. The fully general and invariant
formula of Sachs and Wolfe -- which is reviewed in Section 2
-- allows us to solve for the perturbations with any convenient
choice of gauge and field variables. Although we shall not
work beyond linearized order, it is worth noting that the
result of Sachs and Wolfe can be extended to any desired order
in the weak field expansion. The method is applied for the
generic system of a graviton with a massless, minimally coupled
scalar in Section 3. The scalar and tensor power spectra
are derived in Sections 4 and 5 respectively. In both cases
improved estimates are obtained. Our conclusions comprise
Section 6. The basics of the evolution dependent improvement 
factors have been summarized in the Appendix. Another Appendix
describes a model in which the slow roll paradigm completely
breaks down but our methods can still be employed.

\section{The Sachs and Wolfe Effect}

The gravitational field equations are:
\footnote{By $ R_{\mu\nu}$ and $R$ we denote the Ricci tensor 
and Ricci scalar constructed from the spacelike metric tensor 
$g_{\mu\nu}$. Furthermore, an overdot indicates differentiation 
with respect to co-moving time $t$ while an overprime denotes 
differentiation with respect to conformal time $\eta$.}

\begin{equation}
G_{\mu\nu} \equiv 
R_{\mu\nu} - \frac12 g_{\mu\nu} R = 
8 \pi G \; T_{\mu\nu} 
\;\; , \label{eqsmotion}
\end{equation}
in which $G$ is the Newton constant. A spatially homogeneous 
and isotropic universe can be conveniently represented by the 
stress tensor $T_{\mu\nu}$ of a perfect fluid with energy 
density $\rho$, pressure $p$ and 4-velocity $u^{\mu}$:
\begin{equation}
T_{\mu\nu} \; = \;
(\rho + p) \, u_{\mu} \, u_{\nu} \; + \;
p \, g_{\mu\nu}
\;\; , \label{fluidTmn} 
\end{equation}
where $u^{\mu}$ obeys:
\begin{equation}
u^{\mu} \, u^{\nu} \, g_{\mu\nu} \; = \; -1
\;\; . \label{4vel} 
\end{equation}
To account for the observed structures and obtain a more 
realistic cosmological description, deviations from homogeneity 
and isotropy are essential. Without any reference to their 
origin, it is simple to incorporate such departures as linear 
perturbations on the dynamical variables of the system:
\begin{eqnarray}
g_{\mu\nu}(\eta, {\vec x}) & = &
{\bar g}_{\mu\nu}(\eta) \; + \; 
\delta g_{\mu\nu}(\eta, {\vec x}) 
\;\; , \label{deltagmn} \\
\rho(\eta, {\vec x}) & = &
{\bar \rho}(\eta) \; + \; \delta \rho(\eta, {\vec x}) 
\;\; , \label{deltarho} \\
p(\eta, {\vec x}) & = &
{\bar p}(\eta) \; + \; \delta p(\eta, {\vec x}) 
\;\; . \label{deltap} 
\end{eqnarray}
The unperturbed metric field ${\bar g}_{\mu\nu}$ belongs to 
the Robertson-Walker class of spacetimes and is, therefore, 
conformally flat and characterized by scale factor $a$:
\begin{equation}
{\bar g}_{\mu\nu} \; = \;
a^2(\eta) \, \eta_{\mu\nu} 
\;\; . \label{gmnbar} 
\end{equation}
The unperturbed ${\bar \rho}$ and ${\bar p}$ correspond to the 
average energy density and pressure of the physical system 
respectively. The arbitrariness in the choice of coordinates 
is resolved by employing a frame that moves with the fluid:
\begin{eqnarray}
g_{00} (\eta, {\vec x}) & = &
-1 
\;\; , \label{gauge1} \\
G^{\, i}_{\;\, 0} (\eta, {\vec x}) & = &
0
\; \; \Longleftrightarrow \;\;
u^{\mu} (\eta, {\vec x}) \; = \;
a^{-1}(\eta) \; \delta^{\mu}_{\;\; 0}
\;\; . \label{gauge2} 
\end{eqnarray}
It is important to note the relation between co-moving and 
conformal time intervals:
\begin{equation}
dt \; = \; a(\eta) \; d\eta
\;\; . \label{confcom} 
\end{equation}

\begin{figure}
\centerline{\epsfig{file=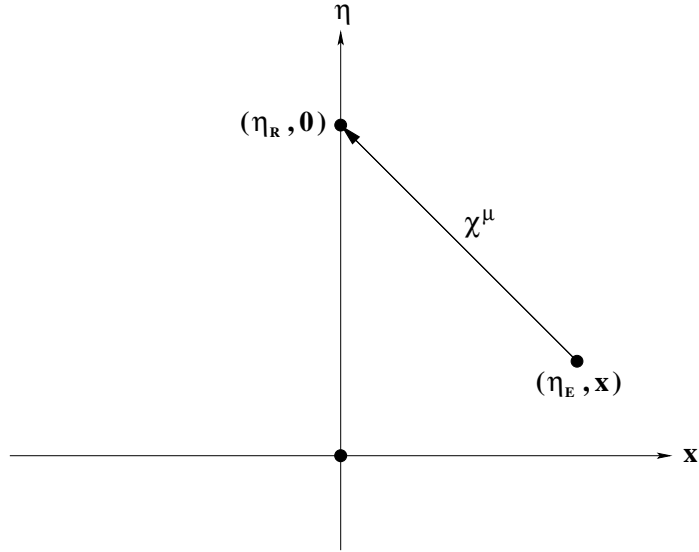,height=2.9in}}
\caption{\footnotesize The light emission 
$(\eta_E \, , \, {\vec x})$ and reception 
$(\eta_R \, , \, {\vec 0})$ events, and the lightlike
\break \mbox{} \hspace{1.9cm} 
geodesic $\chi^{\mu}$.}
\end{figure}

Sachs and Wolfe have computed \cite{SW} the redshift accumulated
by a light ray as it travels in the presence of (\ref{deltagmn}) 
from its emission to its reception {\it (see Figure 2)}. The 
result is quite general as the only relevant ingredient is the 
metric field perturbation:
\begin{equation}
\delta g_{\mu\nu} \; \equiv \;
a^2(\eta) \; h_{\mu\nu}(\eta, {\vec x}) 
\;\; . \label{hmn} 
\end{equation}
If the light signal is observed from direction ${\widehat e}$, 
the wavelength shift $z({\widehat e})$ is given by:
\begin{equation}
1 \, + \, z({\widehat e}) \; = \;
\frac{\left[ \, u^{\mu} \, k_{\mu} \, \right]_{E}}
{\left[ \, u^{\mu} \, k_{\mu} \, \right]_{R}}
\;\; , \label{1+z} 
\end{equation}
where $E$ and $R$ stand for the emission and reception events
respectively, and where $k^{\mu}$ is the 4-momentum of the 
light ray:
\begin{equation}
k^{\mu} (\tau) \; = \;
{\dot \chi}^{\mu} (\tau)
\;\; . \label{4mom} 
\end{equation}
The lightlike geodesic ${\chi}^{\mu}$ satisfies:
\begin{eqnarray}
{\ddot \chi}^{\mu} (\tau) \, + \,
\Gamma^{\mu}_{\;\; \alpha\beta} [\chi(\tau)] \;
{\dot \chi}^{\alpha} (\tau) \;
{\dot \chi}^{\beta} (\tau) & = & 0
\;\; , \label{eqngeodesic} \\
g_{\alpha\beta} [\chi(\tau)] \;
{\dot \chi}^{\alpha} (\tau) \;
{\dot \chi}^{\beta} (\tau) & = & 0
\;\; , \label{eqnlight} 
\end{eqnarray}
and (\ref{eqngeodesic}-\ref{eqnlight}) can be integrated 
to give the following result for $z({\widehat e})$ to first 
order in the perturbation $h_{\mu\nu}$ \cite{SW}:
\begin{eqnarray}
1 \, + \, z({\widehat e})  & = &
\frac{a(\eta_R)}{a(\eta_E)} \, \times 
\label{z1storder} \\ 
& \mbox{} & \hspace{-0.65cm}
\left\{ \, 1 \, - \, \int_{0}^{\eta_R - \eta_E} d\sigma
\left[ \, {\widehat e}^{\, i} \; 
h_{0i \, , \, 0} (x)  \, - \, 
\frac12 \, {\widehat e}^{\, i} \, {\widehat e}^{\, j} \; 
h_{ij \, , \, 0} (x) 
\, \right]_{x^{\mu} \, = \, 
( \, \eta_R - \sigma , \, \sigma {\widehat e} \, )}  
\, \right\}
\, . \nonumber 
\end{eqnarray}

Suppose that thermal radiation of average temperature $T_E$
was emitted from a spacelike surface at the time $\eta_E$ 
of the coordinate system. Then, at the reception event 
$(\eta_R, {\vec 0})$:
\begin{equation}
T_R ({\widehat e}) \; = \;
\frac{T_E ({\widehat e})}{1 + z({\widehat e})}
\;\; , \label{temp}
\end{equation}
so that the first order temperature fluctuation observed
from direction ${\widehat e}$ is:
\begin{eqnarray}
\frac{\Delta T_R}{T_R} ({\widehat e}) & = &
\frac{\Delta T_E}{T_E} ({\widehat e}) \, + 
\label{T1storder} \\ 
& \mbox{} & \hspace{-0.4cm}
\int_{t_E}^{t_R} dt \;
\left[ \, {\widehat e}^{\, i} \; 
h_{0i \, , \, t \,} (x)  \, - \, 
\frac12 \, {\widehat e}^{\, i} \, {\widehat e}^{\, j} \; 
h_{ij \, , \, t \,} (x) 
\, \right]_{x^{\mu} \, = \, \left( \,
t \, , \; {\widehat e} \int_{t}^{t_R} \, dt' \; a^{-1}(t') 
\, \right)}  
\;\; , \nonumber
\end{eqnarray}
and has been expressed in terms of the co-moving time $t$.
Since the accumulated wavelength shift and the temperature
fluctuation are observables, expressions (\ref{z1storder})
and (\ref{T1storder}) are manifestly gauge invariant. 

\begin{figure}
\centerline{\epsfig{file=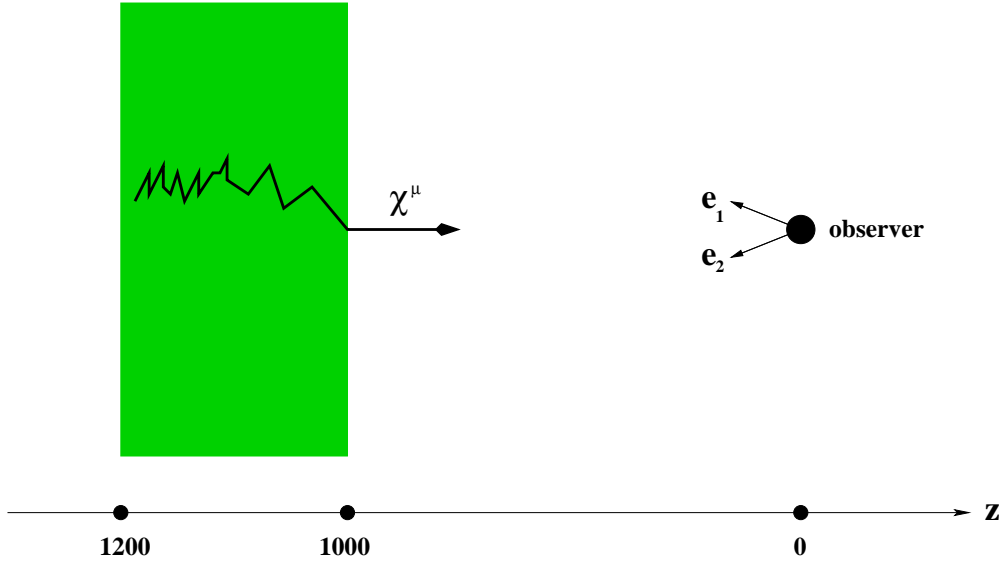,height=2.9in}}
\caption{\footnotesize A typical lightlike geodesic 
$\chi^{\mu}$ on its way to the observer. The graph is not
\break \mbox{} \hspace{1.8cm} 
properly scaled.}
\end{figure}

The Cosmic Microwave Background Radiation (CMBR) consists 
of photons emitted during the period of decoupling. 
\footnote{In the history of the universe, the period of 
decoupling is centered around $z \sim 1089$ with a width 
$\Delta z \sim 195$.}
What is measured is the product of temperature fluctuations 
simultaneously observed from two different directions
{\it (see Figure 3)}. Thus, the connection between the 
measured quantity and the quantum mechanical origin of 
these fluctuations comes from the study of:
\begin{equation}
\left\langle \Omega \left\vert \; 
\frac{\Delta T_R}{T_R} ({\widehat e_1}) \;\;
\frac{\Delta T_R}{T_R} ({\widehat e_2}) \; 
\right\vert \Omega \right\rangle 
\;\; , \label{T1T2} 
\end{equation}
for the appropriate vacuum state $\vert \Omega \rangle$.

\section{The Perturbations of the Graviton-Scalar System}

The Lagrangian describing the system of the graviton and
a minimally coupled scalar is:
\begin{equation}
{\cal L} \; = \;
{1 \over 16 \pi G} \, R \, \sqrt{-g} \; - \;
\frac12 \, \partial_{\mu} \varphi \;
\partial_{\nu} \varphi \;
g^{\mu\nu} \sqrt{-g} \; - \;
V(\varphi) \, \sqrt{-g}
\;\; . \label{L}
\end{equation}
Its dynamical variables are the metric field $g_{\mu\nu}$ 
and the scalar $\varphi$. Both are expressed as a background 
plus a quantum field:
\begin{eqnarray}
g_{\mu\nu}(\eta, {\vec x}) & = &
a^2 (\eta) \, \Bigl( \, \eta_{\mu\nu} \; + \;
h_{\mu\nu}(\eta, {\vec x}) \, \Bigr)
\;\; , \label{gmn} \\
& \equiv &
a^2 (\eta) \, \Bigl( \, \eta_{\mu\nu} \; + \;
\kappa \, \psi_{\mu\nu}(\eta, {\vec x}) \, \Bigr)
\;\; , \label{psimn} \\
\varphi(\eta, {\vec x}) & = &
\varphi_0(\eta) \; + \;
\phi(\eta, {\vec x})
\;\; , \label{varphi}
\end{eqnarray}
where $\kappa^2 \equiv 16 \pi G$ is the loop counting
parameter of quantum gravity.

The background Einstein equations are:
\begin{eqnarray}
3 H^2 & = & 8 \pi G \,
\Bigl( \, \frac12 \, \dot{\varphi}_0^2 + V(\varphi_0) 
\, \Bigr) 
\;\; , \label{E00} \\
(-1 + 2 q) H^2 & = & 8 \pi G \, 
\Bigl( \, \frac12 \, \dot{\varphi}_0^2 - V(\varphi_0) 
\, \Bigr) 
\;\; , \label{Eij}
\end{eqnarray}
where an overdot represents differentiation with respect 
to the co-moving time $t$.
\footnote{The relation between co-moving and conformal 
time derivatives is: $\frac{\partial}{\partial t} = 
\frac1{a} \frac{\partial}{\partial \eta}$.}
Although it is traditional to regard the potential as 
known and then infer the scale factor, with our method 
it is more convenient to regard $a(t)$ -- and hence 
$H(t) = \dot{a}/a$ and $q(t) = -a\ddot{a}/\dot{a}^2$ 
-- as a known function from which the background scalar 
and the potential can be expressed as follows:
\begin{eqnarray}
\dot{\varphi}_0 & = & 
-\sqrt{1 + q(t)} \;\; \frac{H(t)}{\sqrt{4\pi G}} 
\;\; , \label{phidot} \\
V(\varphi_0) & = & 
\Bigl(2 - q(t)\Bigr) \; \frac{H^2(t)}{8 \pi G} 
\;\; . \label{V}
\end{eqnarray}
We parameterize the third derivative of $a(t)$ using 
the variable:
\begin{equation}
r(t) \; \equiv \; \frac1{H(t)} \;
\frac{d}{dt} \ln\Bigl(\sqrt{1 + q(t)} \,\Bigr) 
\;\; . \label{rdef}
\end{equation}
Hence, the derivative of the potential is:
\begin{equation}
V'(\varphi_0) \; = \; 
\Bigl( \, 2 - q(t) + r(t) \, \Bigr) \; \sqrt{1 + q(t)} 
\; \frac{H^2(t)}{\sqrt{4 \pi G}} 
\;\; . \label{V'}
\end{equation}
Higher derivatives of the potential can obviously be 
obtained by taking higher time derivatives of the scale 
factor, for example:
\begin{equation}
V''(\varphi_0) \; = \; 
\Bigg( \, 3 q(t) \, r(t) - r^2(t) - 
\frac{\dot{r}(t)}{H(t)} \, \Bigg) \; H^2(t) 
\;\; . \label{V''}
\end{equation}

A convenient diagonalization of the linearized system
is given in \cite{nctrpw1, rpw1} and is summarized in 
\cite{nctrpw3}. By employing a generalized de Donder gauge
condition:
\begin{equation}
F_{\mu} \; \equiv \;
a \left[ \, 
\psi^{\nu}_{~\mu \, , \, \nu} \, - \, 
\frac12 \, \psi^{\nu}_{~\nu \, , \, \mu} \, - \, 
2 \, a H \, \psi_{\mu 0} \, + \,
2 \delta^0_{~\mu} \, {\sqrt {1 + q}} \; a H \, \phi 
\, \right] \; = \; 0 
\;\; , \label{gauge}
\end{equation}
all linearized fields can be expressed in terms of:
\begin{eqnarray}
\psi^{TT}_{ij}(\eta, {\vec x}) & = &
\int \frac{d^3k}{(2\pi)^3} \; \sum_s \left\{ \,
\epsilon_{ij}({\vec k}, s) \; U_A(\eta, k) \;
e^{i {\vec k} \cdot {\vec x}} \;
\alpha({\vec k}, s) \; + \;
(c.c.) \, \right\}
\; , \qquad \label{psiTT} \\
\psi_{00}(\eta, {\vec x}) &  = &
\int \frac{d^3k}{(2\pi)^3} \; 
\left\{ \, U_C (\eta, k) \;
e^{i {\vec k} \cdot {\vec x}} \; Y({\vec k}) \; + \;
(c.c.) \, \right\}
\; , \label{psi00} 
\end{eqnarray}
as follows:
\begin{eqnarray}
\psi_{0i}(\eta, {\vec x}) & = & 0
\;\; , \label{psi0i} \\
\psi_{ij}(\eta, {\vec x}) & = &
\delta_{ij} \; \psi_{00}(\eta, {\vec x}) 
\; + \; \psi^{TT}_{ij}(\eta, {\vec x}) 
\;\; , \label{psiij} \\
\phi (\eta, {\vec x}) & = &
\frac{1}{\sqrt{1 + q(t)}} \;\;
\frac{1}{H(t) \, a(t)} \;\;
\frac{\partial}{\partial t}
\Bigl( \, a(t) \; \psi_{00}(\eta, {\vec x}) \, \Bigr)
\;\; , \label{phi} 
\end{eqnarray}
The mode functions $U_{A, C}$ are of the form:
\begin{eqnarray}
U_A (\eta, k) & \equiv &
\frac{\sqrt{2}}{a(t)} \; Q_A (\eta, k) 
\;\; , \label{U_A} \\
U_C (\eta, k) & \equiv &
- \, \sqrt{1 + q(t)} \; H(t) \;
\frac1{k} \, Q_C (\eta, k) 
\;\; , \label{U_C} 
\end{eqnarray}
where $Q_{A, C}$ obey \cite{nctrpw3}:
\footnote{While the general solutions to (\ref{Q}-\ref{theta}) 
are known \cite{nctrpw3}, it is only in a particular limit 
that we shall need them for the purposes of this paper.}
\begin{eqnarray}
& \mbox{} &
Q^{''}_{A, C} \; + \; 
\left[ \, k^2 \, - \, 
\frac{\theta^{''}_{A, C}}{\theta_{A, C}} 
\, \right] \, Q_{A, C} = 0
\;\; , \label{Q} \\
& \mbox{} &
\theta_A \; \equiv \; a
\quad , \quad
\theta_C \; \equiv \; 
\frac1{a \, \sqrt{1 + q}}
\;\; . \label{theta} 
\end{eqnarray}
The graviton and scalar creation and annihilation operators 
are canonically normalized:
\begin{eqnarray}
\left[ \, \alpha({\vec k}, s) \; , \; 
\alpha^{\dagger}({\vec k}', s') \, \right]  & = &
(2\pi)^3 \; \delta^3({\vec k} \, - \, {\vec k}') \; 
\delta_{ss'}
\;\; , \label{ccrhmn} \\
\left[ \, Y({\vec k}) \; , \; 
Y^{\dagger}({\vec k}') \, \right] & = &
(2\pi)^3 \; \delta^3({\vec k} \, - \, {\vec k}') 
\;\; . \label{ccrphi} 
\end{eqnarray}
The graviton polarization tensor is purely spatial, transverse
and traceless:
\begin{equation}
\epsilon_{0\mu}({\vec k}, s) \; = \;
k_i \, \epsilon_{ij}({\vec k}, s) \; = \;
\epsilon_{ii}({\vec k}, s) \; = \; 0
\;\; . \label{emn}
\end{equation}
Moreover, summing products of two polarization tensors gives:
\begin{eqnarray}
\sum_s \epsilon_{ij}({\vec k}, s) \;
\epsilon^*_{mn}(\vec{k},s) & = & 
\frac12 \; \Bigl[ \,
\Pi_{im} \, \Pi_{jn} \, + \, 
\Pi_{in} \, \Pi_{jm} \, - \, 
\Pi_{ij} \, \Pi_{mn} \, \Bigr] 
\;\; , \label{emnsum} \\
\Pi_{ij} & \equiv & 
\delta_{ij} \, - \,
\widehat{k}_i \, \widehat{k}_j
\;\; , \label{Pij}
\end{eqnarray}
where $\Pi_{ij}$ is the transverse projector.

In order to apply the basic formula (\ref{T1storder}) for
the first order temperature fluctuation, we must transform
the linearized fields (\ref{psi00}-\ref{phi}) from obeying
the gauge condition (\ref{gauge}) to satisfying the co-moving 
gauge conditions (\ref{gauge1}-\ref{gauge2}). This is 
achieved by effecting the field-dependent coordinate 
transformation:
\begin{equation}
x^{\mu} (x') \; \equiv \;
x'^{\, \mu} \, + \, \kappa \; \varepsilon^{\mu} (x')
\;\; , \label{eps}
\end{equation}
which imposes the co-moving gauge conditions:
\begin{eqnarray}
\varepsilon_0 (\eta, {\vec x}) & = &
- \, \frac{1}{2 a(\eta)} \,
\int_{\eta_E}^{\eta} d\eta' \; a(\eta') \;
\psi_{00}(\eta', {\vec x})
\;\; , \label{eps0} \\
\varepsilon_i (\eta, {\vec x}) & = &
\int_{\eta_E}^{\eta} d\eta' \, \left[ \,
\psi_{0i}(\eta', {\vec x}) \, + \,
\frac{\phi_{, \, i \,} (\eta', {\vec x})}
{\kappa \, \varphi_0 ' (\eta')} \, \right]
\;\; , \label{epsi} 
\end{eqnarray}
on the linearized fields. In particular, since under any
infinitesimal coordinate transformation (\ref{eps}) the 
graviton field transforms to: 
\begin{equation}
{\widetilde \psi_{\mu\nu}} \; = \;
\psi_{\mu\nu} \, + \, 
2 \, \varepsilon_{(\mu \, , \, \nu)} \, - \,
2 H a \, \varepsilon_0 \, \eta_{\mu\nu}
\;\; , \label{psi'}
\end{equation}
for the specific choice (\ref{eps0}-\ref{epsi}) we obtain:
\begin{eqnarray}
{\widetilde \psi_{00}} (\eta, {\vec x}) & = & 0
\;\; , \label{psi00sw} \\
{\widetilde \psi_{0i}} (\eta, {\vec x}) & = & 
- \, \frac{1}{2 a(\eta)} \,
\int_{\eta_E}^{\eta} d\eta' \; a(\eta') \;
\psi_{00 \, , \, i} (\eta', {\vec x}) \, - \,
\frac{\phi_{, \, i \,} (\eta, {\vec x})}
{\kappa \, \varphi_0 ' (\eta)}
\;\; , \label{psi0isw} \\
{\widetilde \psi_{ij}} (\eta, {\vec x}) & = & 
\psi^{TT}_{ij} (\eta, {\vec x}) \, + \,
\delta_{ij} \left[ \,
\psi_{00} (\eta, {\vec x}) \, + \,
H(\eta) \int_{\eta_E}^{\eta} d\eta' \;
a(\eta') \; \psi_{00} (\eta', {\vec x}) 
\, \right] 
\quad \nonumber \\
& \mbox{} & 
- \, 2 \int_{\eta_E}^{\eta} d\eta' \;
\frac{\phi_{, \, ij \,} (\eta', {\vec x})}
{\kappa \, \varphi_0 ' (\eta')}
\;\; . \label{psiijsw} 
\end{eqnarray}

Thus, by construction the graviton field 
(\ref{psi00sw}-\ref{psiijsw}) obeys 
(\ref{gauge1}-\ref{gauge2}). Keeping in mind the definition 
(\ref{psimn}), the first order temperature fluctuation 
(\ref{T1storder}) becomes:
\begin{eqnarray}
\frac{\Delta T_R}{T_R} ({\widehat e}) & = &
\frac{\Delta T_E}{T_E} ({\widehat e}) \, + 
\label{T1psi1} \\ 
& \mbox{} & \hspace{-0.2cm}
\kappa \int_{\eta_E}^{\eta_R} d\eta' \;
\left[ \, {\widehat e}^{\, i} \; 
{\widetilde \psi}_{0i \, , \, 0} (x)  \, - \, 
\frac12 \, {\widehat e}^{\, i} \, {\widehat e}^{\, j} \; 
{\widetilde \psi}_{ij \, , \, 0} (x) 
\, \right]_{x^{\mu} \, = \, \left( \,
\eta' \, , \; ( \eta_R \, - \, \eta' ) \; 
{\widehat e} \; \right)}  
\;\; . \nonumber
\end{eqnarray}
Further reduction of (\ref{T1psi1}) uses:
\begin{eqnarray}
{\widehat e}^{\, i} \; 
{\widetilde \psi}_{0i \, , \, 0} 
(\eta, {\vec x})  & = &
{\widehat e}^{\, i} \, \Biggl[ \, 
- \frac12 \, \psi_{00 \, , \, i} (\eta, {\vec x})  \, + \,
\frac{H(\eta)}{2} \int_{\eta_E}^{\eta} d\eta' \;
a(\eta') \; \psi_{00 \, , \, i} (\eta', {\vec x})  
\nonumber \\
& \mbox{} & \hspace{0.7cm}
- \, \left( \, 
\frac{\phi (\eta, {\vec x})}{\kappa \, {\varphi_0}' (\eta)} 
\, \right)_{\hspace{-0.1cm} , \, 0i} \; \Biggr] 
\;\; , \label{ident1} 
\end{eqnarray}
together with:
\begin{eqnarray}
\hspace{-1.1cm}
- \, \frac12 \, 
{\widehat e}^{\, i} \, {\widehat e}^{\, j} \; 
{\widetilde \psi}_{ij \, , \, 0} 
(\eta, {\vec x})  & = &
{\widehat e}^{\, i} \, {\widehat e}^{\, j} \, 
\left[ \, 
- \, \frac12 \, \psi^{TT}_{ij \, , \, 0} (\eta, {\vec x})  
\, + \, \left( \, 
\frac{\phi (\eta, {\vec x})}{\kappa \, {\varphi_0}' (\eta)} 
\, \right)_{\hspace{-0.1cm} , \, ij} \; \right] 
\nonumber \\
& \mbox{} & \hspace{1.3cm}
- \, \frac12 \Biggl[ \,
\psi_{00 \, , \, 0} (\eta, {\vec x}) \, + \,
\frac{a'(\eta)}{a(\eta)} \; \psi_{00} (\eta, {\vec x}) 
\nonumber \\
& \mbox{} & \hspace{2.2cm}
+ \, 
H'(\eta) \int_{\eta_E}^{\eta} d\eta' \;
a(\eta') \; \psi_{00} (\eta', {\vec x})  
\, \Biggr]
\; . \label{ident2}
\end{eqnarray}
A straightforward computation leads to the final form for
the first order temperature fluctuations:
\begin{eqnarray}
\frac{\Delta T_R}{T_R} ({\widehat e}) & = &
\frac{\Delta T_E}{T_E} ({\widehat e}) \; + 
\label{T1psi2} \\ 
& \mbox{} & 
\frac{\kappa}2 \left[ \,
\psi_{00} (\eta_R, {\vec 0} \hspace{0.05cm} ) \, - \,
H_R \int_{\eta_E}^{\eta_R} d\eta' \;
a(\eta') \; \psi_{00} (\eta', {\vec 0} \hspace{0.05cm} )  
\, \right]
\nonumber \\
& \mbox{} & 
- \; \frac{{\widehat e}^{\, i} \;
\phi_{, \, i \,} (\eta_R, {\vec 0} \hspace{0.05cm} )}
{{\varphi_0}' (\eta_R)} \; + \;
\frac{{\widehat e}^{\, i} \; 
\phi_{, \, i \,} 
( \, \eta_E \, , \, (\eta_R - \eta_E) \, {\widehat e} \, )}
{{\varphi_0}' (\eta_R)} 
\nonumber \\
& \mbox{} & 
\nonumber \\
& \mbox{} & 
- \; \frac{\kappa}2 \; \psi_{00} 
( \, \eta_E \, , \, (\eta_R - \eta_E) \, {\widehat e} \, )
\; - \; \kappa \int_{\eta_E}^{\eta_R} d\eta' \;
\psi_{00 \, , \, 0} 
( \, \eta' \, , \, (\eta_R - \eta') \, {\widehat e} \, )
\nonumber \\
& \mbox{} & 
- \; \frac{\kappa}2 \;
{\widehat e}^{\, i} \, {\widehat e}^{\, j}
\int_{\eta_E}^{\eta_R} d\eta' \;
\psi^{TT}_{ij \, , \, 0} 
( \, \eta' \, , \, (\eta_R - \eta') \, {\widehat e} \, )
\;\; . \nonumber
\end{eqnarray}
The right hand side of (\ref{T1psi2}) consists of a part
associated with the temperature fluctuations of the
emitting surface plus seven terms. The first two of the
latter have no angular dependence and belong to the
monopole contribution. The third term is the dipole
contribution while the fourth is the Sachs-Wolfe velocity 
potential term. The spectra of scalar and tensor 
perturbations that are usually reported reside in the 
fifth (the Sachs-Wolfe potential term) and seventh
terms respectively. The remaining sixth term is sometimes
called the integrated Sachs-Wolfe effect.

\section{The Tensor Power Spectrum}

In 1979, Starobinski\u{\i} \cite{starobinsky3} became 
the first to calculate the tensor power spectrum from 
what would later be called a model of inflation. Subsequent 
computations were in 1982 made by Rubakov, Sazhim and 
Veryaskin \cite{rubakov} and by Fabbri and Pollock 
\cite{fabbri}. The definitive result was obtained by 
Starobinski\u{\i} in 1985 \cite{starobinsky2}. These 
calculations all depend upon a normalization for the 
late-time mode functions whose precise determination 
is our only improvement. However, we shall also carry 
out the computation in a slightly different fashion.

The part of (\ref{T1psi2}) relevant to tensor perturbations
is:
\begin{equation}
\frac{\Delta T_R}{T_R} ({\widehat e}) \,
\Big\vert_{\, h} \; = \;
- \, \frac{\kappa}2 \;
{\widehat e}^{\, i} \, {\widehat e}^{\, j}
\int_{\eta_E}^{\eta_R} d\eta' \;
\psi^{TT}_{ij \, , \, 0} 
( \, \eta' \, , \, (\eta_R - \eta') \, {\widehat e} \, )
\;\; , \label{T1hmn1} 
\end{equation}
and can be expressed as a sum over graviton momenta and 
polarizations:
\begin{equation}
\frac{\Delta T_R}{T_R} ({\widehat e}) \,
\Big\vert_{\, h} \; = \;
\int \frac{d^3k}{(2\pi)^3} \; 
\sum_s \left\{ \,
h({\widehat e}, {\vec k}) \;\;
{\widehat e}^{\, i} \, {\widehat e}^{\, j} \, 
\epsilon_{ij}(\vec{k},s) \;\,
\alpha(\vec{k},s) \; + \;
(c.c.) \, \right\}
\; , \label{T1hmn2}
\end{equation}
where the scalar response function is:
\begin{equation}
h({\widehat e}, {\vec k}) \; = \;
- \frac{\kappa}{2} \,
\int_{t_E}^{t_R} dt \;
\left( \, \frac{\partial}{\partial t} \;
U_A (t, k) \, \right) \;
\exp \left[ \, i {\vec k} \cdot 
{\widehat e} \int_{t}^{t_R} dt' \; a^{-1}(t') \, \right]
\;\; . \label{srf1}
\end{equation}
It is straightforward to compute the expectation value 
(\ref{T1T2}) in the presence of the state which was empty
of gravitons in the distant past:
\begin{equation}
\alpha({\vec k}, s) \, \vert\Omega \rangle \; = \; 0
\;\; , \label{vach} 
\end{equation}
and obtain:
\begin{eqnarray}
\left\langle \Omega \left\vert \; 
\frac{\Delta T_R}{T_R} ({\widehat e_1}) \;\;
\frac{\Delta T_R}{T_R} ({\widehat e_2}) \; 
\right\vert \Omega \right\rangle_{\hspace{-0.1cm} h} 
& = &
\int \frac{d^3k}{(2\pi)^3} \;\;
h({\widehat e}_1, {\vec k}) \;\;
h^{*}({\widehat e}_2, {\vec k}) 
\label{T1T2hmn1} \\
& \mbox{} & \hspace{0.2cm} \times \; 
{\widehat e}^{\, i}_1 \, {\widehat e}^{\, j}_1 \, 
{\widehat e}^{\, m}_2 \, {\widehat e}^{\, n}_2 \, 
\left[ \, \Pi_{im} \, \Pi_{jn} \, - \, 
\frac12 \Pi_{ij} \, \Pi_{mn} \, \right] 
\; . \nonumber 
\end{eqnarray}

The scalar response function (\ref{srf1}) can be explicitly
evaluated because the physical process occurs entirely during 
the epoch of matter domination. If we assume that the onset
of matter domination occurred at a time $t_M$, when the Hubble 
parameter and scale factor were $H_M$ and $a_M$ respectively, 
then at later times:
\footnote{During matter domination, the deceleration parameter 
$q(t)$ is quite well approximated by the constant 
$q_m = +\frac12$.} 
\begin{eqnarray}
Matter 
\quad \Longrightarrow \quad
H(t) & = & 
\frac{H_M}{1 \, + \, \frac32 H_M \, (t - t_M)} 
\;\; , \label{Hmatter} \\
a(t) & = & 
a_M \left[ \,
1 \, + \, \frac32 H_M \, (t - t_M)
\, \right]^{\frac23} 
\;\; . \label{amatter}
\end{eqnarray}
In view of (\ref{Hmatter}-\ref{amatter}), the dimensionless 
variable (\ref{xdef}) equals: 
\begin{equation}
Matter 
\quad \Longrightarrow \quad
x(t,k) \; = \; 
x(t_M, k) \, \left[ \,
1 \, + \, \frac32 H_M \, (t - t_M)
\, \right]^{\frac13} 
\;\; , \label{xmatter}
\end{equation}
In terms of $x$, the radial component of a lightlike geodesic 
times $k$ takes the form:
\begin{equation}
Matter
\quad \Longrightarrow \quad
k \, \int_t^{t_R} \; \frac{dt'}{a(t')} \; = \;
2 x(t_R, k) \, - \, 2 x(t, k) 
\;\; , \label{lgmatter}
\end{equation}
and the scalar response function (\ref{srf1}) becomes:
\footnote{Henceforth in this section, all quantities refer 
to the form they take for a matter dominated universe.}
\begin{equation}
h(\widehat{e}, {\vec k}) \; = \;
- \frac{\kappa}{2} \;
\int_{x_E}^{x_R} \; dx \;
\left( \, \frac{\partial}{\partial x} \;
U_A (t, k) \, \right) \;
e^{2i \, {\widehat k} \, \cdot \, 
{\widehat e} \, (x_R - x)}
\;\; . \label{srfmatter1}
\end{equation}

Further progress in the evaluation of the scalar response 
function (\ref{srfmatter1}) requires an explicit form for 
the mode function. Indeed, the source of our improved 
estimate for the graviton power spectrum is our improved 
derivation of the graviton mode functions \cite{nctrpw2}. 
Since the physical process under study involves modes that 
underwent first horizon crossing at $t = t_1$, the relevant 
form of the mode functions for $t > t_1$ is \cite{nctrpw3}:
\footnote{The subscript $1$ in a quantity signifies its
value at first horizon crossing $t = t_1$. }
\begin{equation}
U_A(t, k) \; = \; 
\frac{-iH_1}{\sqrt {k^3}} \;
\frac{\Gamma(1 - \nu) \; J_{-\nu}(- \frac{x}{q})}
{(- \frac{x}{2q})^{-\nu}} \; \times \;
{\cal C}_{1A}(k) \, \times \, 
{\cal C}_{iA}(k)
\;\; . \label{mfA1}
\end{equation}
It consists of three factors, the first of which is the 
time dependent part:
\begin{equation}
\frac{\Gamma(1 - \nu) \; J_{-\nu}(- \frac{x}{q})}
{(- \frac{x}{2q})^{-\nu}}
\Bigg\vert_{\, q \, = \, \frac12 \, , 
\; \nu \, = \, -\frac32} = \; 
3 \, \sqrt{\frac{\pi}2} \;
\frac{J_{\frac32}(2x)}{(2 x)^{\frac32}} 
\;\; . \label{mfpart1}
\end{equation}
This is a standard result.
\footnote{See, for instance, equation (4.29) of \cite{LL}.}
The remaining two factors in (\ref{mfA1}) represent our 
improvement to the normalization of the mode functions; 
${\cal C}_{1A}$ depends upon the state of the system at 
first horizon crossing:
\begin{equation}
{\cal C}_{1A} (k) \; \equiv \; 
\frac{\frac1{\sqrt{\pi}} \;
\Gamma( \, \frac12 - \frac1{q_1} \, )}
{(- \frac1{2q_1})^{- \frac1{q_1}}} 
\;\; . \label{C1A}
\end{equation}
To get a feeling for ${\cal C}_{1A}$, it is important to 
note that for perfect de Sitter inflation it equals one:
\begin{equation}
q_1 \, = \, -1
\quad \Longrightarrow \quad
{\cal C}_{1A} (k) \, = \, 1 
\;\; , \label{dSC1A}
\end{equation}
while for a more realistic situation one finds to first 
order:
\begin{eqnarray}
q_1 (k) & = & -1 \, + \, \Delta q(k)
\quad \Longrightarrow \quad
\nonumber \\
{\cal C}_{1A} (k) & = & 1 \, + \, \Bigl[ \, 
\psi({\scriptstyle \frac32}) \, + \, \ln 2 \, - \, 1 
\, \Bigr] \Delta q(k)
\quad , \quad
\psi(z) \equiv \frac{\Gamma'(z)}{\Gamma(z)}
\;\; . \qquad \label{realC1A}
\end{eqnarray}
The factor ${\cal C}_{iA}$ depends upon previous evolution. 
Had there been no evolution in $q$ from the initial time 
$t_i$ to $t_1$, its value would be one:
\begin{equation}
q(t) \, = \, {\bar q}
\quad \Longrightarrow \quad
{\cal C}_{iA}(k) = 1 
\;\; . \label{apprCiA}
\end{equation}
When -- as is the physical case -- there is a mild evolution,
it results in small deviations about (\ref{apprCiA}) whose
explicit form is given in the first Appendix.

In view of (\ref{mfpart1}), we can express (\ref{mfA1}) with 
its conventional slow-roll normalization times the two 
correction factors:
\begin{equation}
U_A(t, k) \; = \; 
\frac{-iH_1}{\sqrt {k^3}} \;
3 \left[ \, \frac{\sin(2x)}{8x^3} \, - \,
\frac{\cos(2x)}{4x^2} \, \right] \, \times \,
{\cal C}_{1A}(k) \, \times \, 
{\cal C}_{iA}(k)
\;\; . \label{mfA2}
\end{equation}

With the infrared approximation (\ref{mfA2}) it is possible
to exactly evaluate the scalar response function
(\ref{srfmatter1}):
\begin{eqnarray}
h(\widehat{e}, {\vec k})
& = &
\frac{i\kappa H_1}{\sqrt{2k^3}} \;
\frac3{\sqrt{2}} \;
{\cal C}_{1A} \;
{\cal C}_{iA} \; e^{2i \, w \, x_{R}} \;
\Biggl\{ \, \Biggl[ \,
\frac{\sin(2x)}{8x^3} \, - \,
\frac{\cos(2x)}{4x^2}
\label{srfmatter2} \\
& \mbox{} &
- \, i w \left( \, \frac{\sin(2x)}{8x^2} \, - \,
\frac{\cos(2x)}{4x} \, \right) \, - \,
w^2 \; \frac{\sin(2x)}{4x} \; \Biggr] e^{-2i \,w x}
\nonumber \\
& \mbox{} &
+ \, \frac{w}{4} \, (1 - w^2) \Bigl[ \,
{\rm Ei}\Bigl( \, 2 i (1 - w) x \, \Bigr) \, - \,
{\rm Ei}\Bigl( \, -2 i (1 + w) x \, \Bigr)
\, \Bigr] \, \Biggr\}
\, \Biggr\vert_{x_E}^{x_R}
\;\; , \nonumber
\end{eqnarray}
where, to economize on writing, we have defined
$w \equiv \widehat{k} \cdot \widehat{e}$ as the cosine of the
angle between the unit vectors $\widehat{k}$ and $\widehat{e}$.
However, there is no point in retaining the full complexity
of this result. It is easy to check that the term inside the
curly brackets falls like $x^{-2}$ for large $x$. Therefore,
potentially observable effects must derive from modes which
had not yet experienced second horizon crossing at the time
of emission. This implies $x_E \ll 1$. The modes which produce
anisotropies within our current horizon volume must also have
experienced second horizon crossing by the time of reception.
Hence we can also assume $x_R \gg 1$. It follows that the only
significant contribution comes from the lower limit, for which
we may as well take the limiting form relevant to small $x_E$:
\footnote{Although our technique has been different, this result
seems to agree with Starobinski\u{\i}'s equation (12) 
\cite{starobinsky2}.}
\begin{eqnarray}
h(\widehat{e}, {\vec k}) \,
\Bigg\vert_{\, x_E \, \ll \, 1}^{\, x_R \, \gg \, 1} & = &
- \frac{i\kappa H_1}{\sqrt{k^3}} \; 
{\cal C}_{1A} \; {\cal C}_{iA} \; 
e^{2i \, w \, x_{R}} 
\label{srfmatter3} \\
& \mbox{} & 
\times \, 
\left\{ \, \frac12 \, - \, \frac34 w^2 \, - \,
\frac38 \, w \, (1 - w^2) 
\left[ \, \ln \left( \frac{1 + w}{1 - w} \right) \, + \,
i \pi \, \right] \, \right\} 
\;\; . \nonumber
\end{eqnarray}

The angular dependence in our expression (\ref{srfmatter3}) 
for the scalar response function is complicated. However, 
one can recognize some of the factors as spherical harmonics 
with zenith angle $\theta = {\rm arccos}(\widehat{e} \cdot
\widehat{k})$ and azimuthal angle $\phi = 0 \,$:
\begin{eqnarray}
\frac12 \, - \, \frac34 \, w^2 & = & 
\frac{\sqrt{\pi}}2 \, Y_{00} \, - \,
\sqrt{\frac{\pi}5} \, Y_{20} 
\;\; , \label{Ylm1} \\
-\frac38 \, w \, (1 - w^2) & = & 
-\frac32 \, \sqrt{\frac{2\pi}{105}} \, Y_{32} 
\;\; . \label{Ylm2}
\end{eqnarray}
It makes sense to decompose the scalar response function into 
a part depending only upon $ k \equiv \Vert \vec{k} \Vert$ 
and an angular factor $\Theta$, with the $Y_{00}$ term in the 
latter bearing  unit normalization:
\begin{equation}
h(\widehat{e}, {\vec k}) \,
\Bigg\vert_{\, x_E \, \ll \, 1}^{\, x_R \, \gg \, 1} \; = \;
\frac{-i\kappa H_1}{\sqrt{k^3}} \; 
{\cal C}_{1A} \; {\cal C}_{iA} \; 
\frac{\sqrt{\pi}}2 \; 
\Theta(\widehat{e}, \vec{k}) 
\;\; . \label{srfmatter4} 
\end{equation}
Obviously:
\begin{equation}
\Theta(\widehat{e}, \vec{k}) \; \equiv \; 
\frac{e^{2i \, w \, x_{R}}}{\sqrt{4\pi}}  
\left\{ \, 2 - 3w^2 - \frac32 \, w \, (1 - w^2) 
\left[ \, \ln \left( \frac{1 + w}{1 - w} \right) \, + \,
i \pi \, \right] \, \right\} 
\;\; . \label{Theta}
\end{equation}

We define the ``graviton power spectrum'' in terms of the 
radial factor:
\begin{eqnarray}
{\cal P}_h (k) & \equiv & 
\frac{k^3}{4\pi^2} \;
\Bigl\Vert \, \frac{-i\kappa H_1}{\sqrt{k^3}} \;
{\cal C}_{1A} \; {\cal C}_{iA} \; 
\frac{\sqrt{\pi}}2 \, \Bigr\Vert^2 
\;\; , \label{Ph1} \\
& = &  
G H_1^2 (k) \;\; 
{\cal C}^2_{1A}(k) \;\; 
\Vert \, {\cal C}_{iA} (k) \, \Vert^2  
\;\; . \label{Ph2}
\end{eqnarray}
Because the literature abounds with different conventions 
for this quantity, we correspond ${\cal P}_h (k)$ to the 
symbol $\delta_h^2 (k)$ used by Mukhanov, Feldman and 
Brandenberger \cite{MFB}, to the variable ${\cal P}_g (k)$ 
used by Liddle and Lyth \cite{LL}, and to the quantity 
$A_T^2 (k)$ used by Lidsey {\it et al.} \cite{LLK}:
\begin{equation}
{\cal P}_h (k) \; = \;
\frac{9\pi}4 \, \delta_h^2 (k) \; = \;
\frac{\pi}{16} \, {\cal P}_g (k) \; = \;
\frac{25\pi}4 \, A_T^2 (k)
\;\; . \label{Phconv}
\end{equation}
Perhaps the clearest specification of ${\cal P}_h (k)$ is 
to state how it enters the temperature correlation function:
\begin{eqnarray}
\left\langle \Omega \left\vert \; 
\frac{\Delta T_R}{T_R} ({\widehat e_1}) \;\;
\frac{\Delta T_R}{T_R} ({\widehat e_2}) \; 
\right\vert \Omega \right\rangle_{\hspace{-0.1cm} h} 
& = &
2 \int_0^{\infty} \frac{dk}{k} \; 
{\cal P}_h (k) 
\int \frac{d^2 {\widehat k}}{4\pi} \; 
\Theta({\widehat e}_1, \vec{k}) \;\, 
\Theta^{*}({\widehat e}_2, \vec{k}) \; 
\nonumber \\
& \mbox{} & \hspace{-1cm} \times \; 
{\widehat e}^{\, i}_1 \, {\widehat e}^{\, j}_1 \, 
{\widehat e}^{\, m}_2 \, {\widehat e}^{\, n}_2 \, 
\left[ \, \Pi_{im} \, \Pi_{jn} \, - \, 
\frac12 \Pi_{ij} \, \Pi_{mn} \, \right] 
\;\; . \label{T1T2hmn2}
\end{eqnarray}

The leading order slow roll result for ${\cal P}_h(k)$ 
is typically expressed in terms of the value of the scalar 
potential at horizon crossing. Using (\ref{V}) it can be 
converted to our notation:
\begin{equation}
\frac{8 \pi}{3} G^2 V_1 \; = \;
G H^2_1 \; \Bigl( \, \frac{2 - q_1}{3} \, \Bigr) 
\;\; . 
\end{equation}
Our correction factors of $\frac3{2 - q_1}$, $C^2_{1A}(k)$ 
and $\Vert C_{iA}(k) \Vert^2$ are typically near one for 
slow roll inflation. Note especially the factor 
$\Vert C_{iA}(k) \Vert^2$, which represents the effect of 
evolution from the beginning of inflation up to horizon 
crossing, as required by the analysis of Wang, Mukhanov 
and Steinhardt \cite{WMS}.

It is elementary to verify that there is no monopole 
contribution to (\ref{T1T2hmn2}) by fixing one of the 
two directions, for instance ${\widehat e}_2$, and 
integrating over the other:
\begin{equation}
Monopole 
\quad \Longrightarrow \quad
\frac1{4\pi}
\int d^2 {\widehat e}_1 \;
\left\langle \Omega \left\vert \; 
\frac{\Delta T_R}{T_R} ({\widehat e_1}) \;\;
\frac{\Delta T_R}{T_R} ({\widehat e_2}) \; 
\right\vert \Omega \right\rangle_{\hspace{-0.1cm} h} 
\;\; . \label{monopole} 
\end{equation}
If we take the $z$-axis to be along the ${\widehat k}$ 
direction, we can express ${\widehat e_1}$ in terms of 
the zenith angle $\theta$ and azimuthal angle $\phi$:
\begin{equation}
{\widehat e}_1 \; = \;
( \, \sin\theta \, \cos\phi \; , \;
\sin\theta \, \sin\phi \; , \;
\cos\theta \, )
\;\; . \label{e1} 
\end{equation}
The resulting azimuthal integration is simple:
\begin{equation}
\int_0^{2\pi} \frac{d\phi}{2\pi} \;\;
{\widehat e}^{\, i}_1 \, {\widehat e}^{\, j}_1 
\; = \;
\frac12 \, \Pi^{ij} \, \sin^2\theta \,
\; + \; 
{\widehat k}^{\, i} \, {\widehat k}^{\, j} \,
\cos^2\theta
\;\; , \label{eiej} 
\end{equation}
and the properties of $\Pi_{ij}$ ensure that (\ref{eiej})
gives a vanishing monopole contribution (\ref{monopole}).

In a similar fashion, it can be proved that (\ref{T1T2hmn2})
contains no dipole component:
\begin{equation}
Dipole 
\quad \Longrightarrow \quad
\frac1{4\pi}
\int d^2 {\widehat e}_1 \; {\widehat e}^{\, j}_1 \;
\left\langle \Omega \left\vert \; 
\frac{\Delta T_R}{T_R} ({\widehat e_1}) \;\;
\frac{\Delta T_R}{T_R} ({\widehat e_2}) \; 
\right\vert \Omega \right\rangle_{\hspace{-0.1cm} h} 
\; = \; 0
\;\; , \label{dipole} 
\end{equation}
where, as for the monopole case, direction ${\widehat e}_2$
has been fixed.

\section{The Scalar Power Spectrum}

The spectrum of scalar perturbations can be computed from the
Sachs-Wolfe potential term in (\ref{T1psi2}):
\begin{equation}
\frac{\Delta T_R}{T_R} ({\widehat e}) \,
\Big\vert_{\, {\rm SW}} \; = \;
- \; \frac{\kappa}2 \; \psi_{00} 
( \, \eta_E \, , \, (\eta_R - \eta_E) \, {\widehat e} \, )
\;\; . \label{T1phi1} 
\end{equation}
By virtue of (\ref{psi00}) we have:
\begin{equation}
\frac{\Delta T_R}{T_R} ({\widehat e}) \,
\Big\vert_{\, {\rm SW}} \; = \;
- \, \frac{\kappa}2 \,
\int \frac{d^3k}{(2\pi)^3} \; 
\left\{ \, U_C (\eta_E, k) \;
e^{ik \, (\eta_R - \eta_E) \, 
{\widehat k} \, \cdot \, {\widehat e}} \; \;
Y({\vec k}) \; + \; (c.c.) \, \right\}
\;\; . \label{T1phi2} 
\end{equation}
In the presence of the state without any scalars in the 
distant past:
\begin{equation}
Y({\vec k}) \, \vert\Omega \rangle \; = \; 0
\;\; , \label{vacphi} 
\end{equation}
the temperature correlation function (\ref{T1T2}) becomes:
\begin{eqnarray}
\left\langle \Omega \left\vert \; 
\frac{\Delta T_R}{T_R} ({\widehat e_1}) \;\;
\frac{\Delta T_R}{T_R} ({\widehat e_2}) \; 
\right\vert \Omega \right\rangle_{\hspace{-0.05cm} {\rm SW}} 
& = &
\frac{\kappa^2}{4}
\int \frac{d^3k}{(2\pi)^3} \;\;
\Vert \, U_C(\eta_E, k) \, \Vert^2 
\nonumber \\
& \mbox{} & \hspace{1cm} \times \; 
e^{2i \, (x_R - x_E) \, {\widehat k} \, \cdot \,
(\, {\widehat e}_1 \, - \, {\widehat e}_2 \,)}
\;\; , \qquad\qquad \label{T1T2phi1} 
\end{eqnarray}
in terms of the dimensionless variable (\ref{xdef}).

The relevant form of the mode functions is for $t > t_1$
\cite{nctrpw3}:
\begin{eqnarray}
U_C (t, k) \Big\vert_{\, x \, \ll \, 1} & = &
\frac{-H_1}{\sqrt {2k^3 \, (1 + q_1)}} \;
\frac{H(t)}{a(t)} \;
\int_{t_1}^{t} dt' \; a(t') \;
[ \, 1 + q(t') \, ] 
\nonumber \\
& \mbox{} & \hspace{5.3cm}
\times \; {\cal C}^{*}_{1C}(k) \, \times \, 
{\cal C}^{*}_{iC}(k)
\;\; , \qquad \label{mfC1}
\end{eqnarray}
where $t_1$, as always, signals first horizon crossing.
In analogy with Section 4, the normalization factor
${\cal C}_{1C}$ depends upon the state of the system 
at $t_1$. It is expressed in terms of $r(t)$ -- 
defined in (\ref{rdef}) -- and the parameter $q_C(t)$:
\begin{equation}
q_C(t) \equiv - \, \frac{q(t)}{1 + r(t)} \, + \,
\frac{\frac{\dot{r}(t)}{H(t)}}
{\left[ \, 1 + r(t) \, \right]^{\, 2}} \;\; . \label{qC}
\end{equation}
The expression is:
\begin{eqnarray}
{\cal C}_{1C} \; \equiv \;
\frac{ \frac{1}{\sqrt \pi} \, 
\Gamma( \, \frac12 + \frac{1}{q_{1C}} \, ) }
{ \left( \, \frac{1}{2q_{1C}} 
\, \right)^{\frac{1}{q_{1C}}} } \;
e^{-i \frac{\pi}{q_{1C}}} \;
\cos( {\scriptstyle \frac{\pi}{q_{1C}}} ) \;
( \, 1 + r_1 \, )^{\frac{1}{q_{1C}}}
\;\; & , & \label{C1C} \\
q_{1C} \; \equiv \; q_C(t_1)
\quad , \quad
r_1 \; \equiv \; r(t_1)
\;\; & . &
\end{eqnarray}
During inflation -- in fact, quite generally -- the parameter
$r$ is typically zero. Therefore:
\begin{equation}
r(t) \, = \, 0
\quad \Longrightarrow \quad
q_C(t) \, = \, -q(t)
\quad \Longrightarrow \quad
{\cal C}_{1C} (k) \, = \,
{\cal C}_{1A} (k) \;
e^{i \frac{\pi}{q_1}} \;
\cos( {\scriptstyle \frac{\pi}{q_1}} )
\qquad \label{dSC1C}
\end{equation}
More generally, if $r$ is small we can write to first order:
\begin{equation}
q_C(t) \, = \, 
-q(t) \, + \, q(t) \, r(t) \, + \, \frac{\dot{r} (t)}{H(t)}
\;\; . \label{apprqC}
\end{equation}
Consequently, as in (\ref{realC1A}), we have to first order:
\begin{eqnarray}
q_1 (k) & = & -1 \, + \, \Delta q(k)
\quad \Longrightarrow \quad
\nonumber \\
\Vert \, {\cal C}_{1C} (k) \, \Vert & = & 
1 \, + \, \Bigl[ \, \psi({\scriptstyle \frac32}) \, - \, 1 
\, \Bigr] \,
\left[ \, \Delta q(k) \, + \, r_1 \, - \, 
\frac{\dot{r} (t)}{H(t)} \, \right]
\;\; . \label{realC1C}
\end{eqnarray}
The other factor, ${\cal C}_{iC}$, depends upon evolution 
from $t_i$ to $t_1$. Just like ${\cal C}_{iA}$, it equals
one when $q$ is constant; its general form can be found 
in the first Appendix.

Because the physical process takes place entirely during 
pure matter domination:
\footnote{Henceforth in this section, all quantities refer 
to the form they take for a matter dominated universe.}
\begin{equation}
Matter \quad \Longrightarrow \quad
\int_{t_1}^t dt' \; a(t') \,
[ \, 1 + q(t') \, ] \; \sim \;
\frac{a(t) \, [ \, 1 + q(t) \, ]}{H(t)} \; \sim \;
\frac{3 a_E}{2 H_E}
\;\; . \label{mdint}
\end{equation}
Thus, the mode functions can be expressed as a product 
of the conventional slow roll normalization with the two 
correction factors:
\begin{equation}
U_C (t_E, k) \; = \; 
\frac{3H_1}{2 \, \sqrt {2k^3 \, (1 + q_1)}} 
\; \times \; {\cal C}^{*}_{1C}(k) 
\, \times \, {\cal C}^{*}_{iC}(k)
\; , \label{mfC2}
\end{equation}
and the temperature correlation function takes the form:
\begin{eqnarray}
\left\langle \Omega \left\vert \; 
\frac{\Delta T_R}{T_R} ({\widehat e_1}) \;\;
\frac{\Delta T_R}{T_R} ({\widehat e_2}) \; 
\right\vert \Omega \right\rangle_{\hspace{-0.05cm} {\rm SW}} 
& = &
9 \pi G 
\int \frac{d^3k}{(2\pi)^3} \;
\frac{1}{2k^3} \;
\frac{H_1^2}{1 + q_1} \;
\Vert \, {\cal C}_{1C}(k) \, \Vert^2 
\nonumber \\
& \mbox{} & \times \; 
\Vert \, {\cal C}_{iC}(k) \, \Vert^2 \;\;
e^{2i \, (x_R - x_E) \, {\widehat k} \, \cdot \,
(\, {\widehat e}_1 \, - \, {\widehat e}_2 \,)}
\;\;\; \qquad \label{T1T2phi2} 
\end{eqnarray}
The identity:
\begin{equation}
\int \frac{d^3k}{(2\pi)^3} \; f(k) \; 
e^{i {\vec k} \, \cdot \, {\vec x}}
\; = \;
\frac{1}{2\pi^2}
\int_0^{\infty} dk \; k^2 \; f(k) \; 
\frac{\sin x}{x}
\; \; , \label{identity}
\end{equation}
reduces (\ref{T1T2phi2}) to its final form:
\begin{eqnarray}
\left\langle \Omega \left\vert \; 
\frac{\Delta T_R}{T_R} ({\widehat e_1}) \;\;
\frac{\Delta T_R}{T_R} ({\widehat e_2}) \; 
\right\vert \Omega \right\rangle_{\hspace{-0.05cm} {\rm SW}} 
& = &
\frac{9G}{4\pi} \, 
\int_0^{\infty} \frac{dk}{k} \;
\frac{H_1^2}{1 + q_1} \;
\Vert \, {\cal C}_{1C}(k) \;\; {\cal C}_{iC}(k) \, \Vert^2 
\nonumber \\
& \mbox{} & \times \; 
\frac{\sin \Bigl[ \, 2(x_R - x_E) \, \Vert \, 
{\widehat e}_1 \, - \, {\widehat e}_2 \, \Vert \, \Bigr]} 
{2(x_R - x_E) \, \Vert \, 
{\widehat e}_1 \, - \, {\widehat e}_2 \, \Vert}
\; . \qquad\quad \label{T1T2phi3} 
\end{eqnarray}

The ``scalar power spectrum'' is defined by the way it 
enters the correlation function between temperature 
fluctuations observed from directions $\widehat{e}_1$ 
and $\widehat{e}_2$:
\begin{equation}
\left\langle \Omega \left\vert \; 
\frac{\Delta T_R}{T_R} ({\widehat e_1}) \;
\frac{\Delta T_R}{T_R} ({\widehat e_2}) \; 
\right\vert \Omega \right\rangle_{\hspace{-0.05cm} {\rm SW}} 
\! = \! \int_0^{\infty} \! \frac{dk}{k} \,
{\cal P}_{\rm SW}(k) \!
\int \! \frac{d^2 {\widehat k}}{4\pi} \;
e^{2i \, (x_R - x_E) \, {\widehat k} \cdot 
( {\widehat e}_1 - {\widehat e}_2 )}
\qquad \label{T1T2phi} 
\end{equation}
Hence, we obtain:
\begin{equation}
{\cal P}_{\rm SW}(k) \; = \; 
\frac{9}{4\pi} \, 
\frac{G H_1^2}{1 \, + \,  q_1} \;\;
\Vert \, {\cal C}_{1C}(k) \, \Vert^2 \;\; 
\Vert \, {\cal C}_{iC}(k) \, \Vert^2 
\;\; . \label{Pphi}
\end{equation}
We again correspond ${\cal P}_{\rm SW} (k)$ to the symbol 
$\delta (k)$ used by Mukhanov, Feldman and Brandenberger 
\cite{MFB}, to the variable ${\cal P}_{\cal R} (k)$ used 
by Liddle and Lyth \cite{LL}, and to the quantity 
$A_S^2 (k)$ used by Lidsey {\it et al.} \cite{LLK}:
\begin{equation}
{\cal P}_{\rm SW} (k) \; = \;
\frac{25}4 \, \Vert \, \delta (k) \, \Vert^2  \; = \;
\frac94 \, {\cal P}_{\cal R}(k) \; = \;
\frac{225}{16} \, A_S^2 (k) 
\;\; . \label{Pphiconv}
\end{equation}

The leading slow roll result for ${\cal P}_{\rm SW}(k)$ 
is usually expressed in terms of the scalar potential and 
its derivative at the time of horizon crossing. Using 
(\ref{V}) and (\ref{V'}) we can convert this to our notation:
\begin{equation}
96 \pi G^3 \; \frac{V^3_1}{V^{\prime \, 2}_1} \; = \;
\frac{9}{4 \pi} \; \frac{G H_1^2}{1 + q_1} \;\, 
\frac{(2-q_1)^3}{3 (2 - q_1 + r_1)^2} 
\;\; . \label{srpred}
\end{equation}
Our correction factors of 
$\frac{3 (2 - q_1 + r_1)^2}{(2 - q_1)^3}$, 
$\Vert C^2_{1C}(k) \Vert^2$ and $\Vert C_{iC}(k) \Vert^2$ 
are typically near one for slow roll inflation. Consistent 
with the analysis of Wang, Mukhanov and Steinhardt \cite{WMS}, 
there is a factor $\Vert C_{iC}(k) \Vert^2$ which represents 
the effect of evolution from the beginning of inflation up 
to horizon crossing.

\section{Epilogue}

We have taken advantage of a recent, exact solution for the 
mode functions of scalar-driven cosmology \cite{nctrpw3} to 
re-compute the scalar and tensor power spectra for anisotropies 
in the cosmic microwave background. For completeness, and 
to emphasize its inherent gauge invariance, we have also 
reviewed the standard computation of the Sachs-Wolfe effect. 
The principal new feature is our expressions for the 
normalization factors that were built-up during inflation.

We have not expanded the temperature correlation function 
in spherical harmonics. Nonetheless, since our results take
the form of the standard normalization times correction 
factors, it should suffice to simply multiply the standard 
result by these correction factors evaluated at the wavenumber
appropriate for the $l$-th multipole moment:
\begin{equation}
k \; = \;
\frac12 \, l \, a_0 \, H_0
\;\; . \label{lth}
\end{equation}
The tensor correction factors ${\cal C}_{1A}$ and 
${\cal C}_{iA}$ are given by (\ref{C1A}) and (\ref{CiA}) 
respectively; the analogous scalar factors ${\cal C}_{1C}$ 
and ${\cal C}_{iC}$ by (\ref{C1C}) and (\ref{CiC}). 

How observable are the correction factors we have found? 
Since it is likely to require a major effort to detect a 
non-zero tensor amplitude, the fractional improvement we 
give for this probably does not matter. On the other hand,
precision measurements of the scalar amplitude might very 
well be sensitive to the structure we provide. The greatest 
advantage of our formalism is not the incremental improvements 
it offers for the standard, slow roll regime but rather its 
applicability to exotic scenarios that lie beyond the slow
roll paradigm. We present an example in the second appendix.

Finally, we disagree slightly with the standard treatment 
of the tensor contribution. The original authors seem to 
have averaged over graviton polarizations before taking 
the expectation value. This makes a small but possibly 
significant difference in the tensor contribution to the 
multipole moments of the temperature fluctuations 
correlation function.

\newpage

\centerline{\bf Acknowledgements}
 
This work was partially supported by European Union grants 
HPRN-CT-2000-00122 and HPRN-CT-2000-00131, by the DOE contract 
DE-FG02-97ER41029, and by the Institute for Fundamental Theory 
at the University of Florida.

\newpage

\section{Appendix: The evolution dependent correction factors}

A central feature of our exact solutions is the transfer 
matrix, ${\cal M}_I(t,t_i,k)$. There is an $I=A$ transfer 
matrix for the graviton mode function and an $I=C$ one for 
the scalar mode function. Each of them is the time-ordered 
product of the exponential of a line integral:
\begin{eqnarray}
{\cal M}_I(t,t_i,k) & \equiv &
P\left\{ 
\exp\left[ \, \int_{t_i}^t dt' \;
{\cal A}_I(t',k) \, \right] \, \right\} \; ,
\\
& \equiv &
\sum_{n=0}^{\infty} \, \int_{t_i}^t dt_1 
\int_{t_i}^{t_1} dt_2 \, \dots 
\int_{t_i}^{t_{n-1}} dt_n \;
{\cal A}_I(t_1,k) \cdots {\cal A}_I(t_n,k) 
\; . \qquad\;\;
\end{eqnarray}
The exponent matrix ${\cal A}_I(t,k)$ vanishes whenever 
there is no evolution of the appropriate $q_I(t)$:
\footnote{Recall the definition (\ref{rdef}) of the 
parameter $r(t)$.}
\begin{eqnarray}
Graviton \quad \Longrightarrow \quad q_A(t) & = & q(t) 
\;\; , \\
Scalar \quad \Longrightarrow \quad q_C(t) & = & 
- \, \frac{q(t)}{1 + r(t)} \, + \,
\frac{\frac{\dot{r}(t)}{H(t)}}
{\left[ \, 1 + r(t) \, \right]^{\, 2}} 
\;\; .
\end{eqnarray}
There is a similar dichotomy for the appropriate physical 
wave number expressed in Hubble units,
\begin{eqnarray}
Graviton \quad \Longrightarrow \quad x_A(t,k) & = & 
\frac{k}{a(t) H(t)} \;\; , \\
Scalar \quad \Longrightarrow \quad x_C(t,k) & = & 
\frac{x_A(t,k)}{1 + r(t)} \;\; .
\end{eqnarray}
With these definitions the exponent matrix takes the form:
\begin{equation}
{\cal A}_I(t,k) \; = \; 
\frac{\pi}4 \, \dot{\nu}_I 
\left( \,
\matrix{ 
{\rm csc}(\nu_I \, \pi) \; c_{\nu_I}(- \frac{x_I}{q_I}) 
\; & \; 
- 2 i \, d_{\nu_I}(- \frac{x_I}{q_I}) \cr 
- 2 i \, {\rm csc}^2 \, (\nu_I \pi) \; 
b_{\nu_I}(- \frac{x_I}{q_I}) 
\; & \; 
-{\rm csc}(\nu_I \, \pi) \; c_{\nu_I}(- \frac{x_I}{q_I}) }
\, \right) .
\end{equation}
where the various coefficient functions are:
\begin{eqnarray}
b_{\nu}(z) & = & 
{1 \over 2 \sqrt{\pi}} \; 
\sum_{n=1}^{\infty} \,
{(-1)^n \; \Gamma(n - \nu - \frac12) \;
z^{2n - 2\nu} \; (n - \nu)^{- 1} 
\over 
\Gamma(n) \; \Gamma(n - \nu + 1) \; \Gamma(n - 2\nu + 1) } 
\;\; , \\
c_{\nu}(z) & = & 
- {4 \over \pi} \, \sin(\nu \pi) \Bigl[ \, 
\psi(\nu) - 1 - \ln({\scriptstyle \frac12} z) \, \Bigr] 
\nonumber \\
& \mbox{} & 
- {1 \over \sqrt{\pi}} \; 
\sum_{n=1}^{\infty} \, 
{(-1)^n \; \Gamma(n - \frac12) \; z^{2n} \; n^{-1} 
\over 
\Gamma(n + \nu) \; \Gamma(n + 1) \; \Gamma(n - \nu + 1) } 
\;\; , \\
d_{\nu}(z) & = & 
{1 \over 2 \sqrt{\pi}} \;
\sum_{n=0}^{\infty} \,
{(-1)^n \; \Gamma(n + \nu - \frac12) \; 
z^{2n + 2\nu} \; (n + \nu)^{-1} 
\over 
\Gamma(n + 2\nu) \; \Gamma(n + \nu + 1) \; \Gamma(n + 1)} 
\;\; ,
\end{eqnarray}
and we have defined:
\begin{equation}
\nu_I(t) \; \equiv \; 
\frac12 \, - \, q_I^{-1}(t)
\quad , \quad
\psi(z) \; \equiv \; 
\frac{\Gamma'(z)}{\Gamma(z)}
\;\; . 
\end{equation}

We can now give precise definitions for the evolution dependent 
normalization factors:
\begin{eqnarray}
{\cal C}_{iA}(k) & \equiv & 
{\cal M}^{11}_A(t_1,t_i,k) \; + \;
{\cal M}^{12}_A(t_1,t_i,k) \; 
e^{i \frac{\pi}{q_i}} \;
{\rm sec}({\scriptstyle \frac{\pi}{q_i}})
\;\; , \label{CiA} \\
{\cal C}_{iC}(k) & \equiv & 
{\cal M}^{21}_C(t_1,t_i,k) \; + \;
{\cal M}^{22}_C(t_1,t_i,k) \;
e^{-i \frac{\pi}{q_{iC}}} \;
{\rm sec}({\scriptstyle \frac{\pi}{q_{iC}}}) 
\;\; . \label{CiC}
\end{eqnarray}
The subscript $i$ denotes the initial value of the respective
parameter. Since during inflation one typically has:
\begin{equation}
\dot{\nu}(t) \; = \;
\frac{\dot{q}(t)}{q^2(t)} 
\; \ll \; 1
\;\; , 
\end{equation}
it ought to be a very good approximation to simply take the 
first several terms of the series expansion of the transfer 
matrix in estimating these corrections:
\begin{eqnarray}
{\cal M}_I^{11} & \sim & 
1 \, + \, \int_{t_i}^{t_1} dt \, \gamma_I(t) \, + \, 
\int_{t_i}^{t_1} dt \int_{t_i}^t dt' 
\Bigl[ \, \gamma_I(t) \, \gamma_I(t') - 
\delta_I(t) \, \beta_I(t') \, \Bigr] 
\; , \qquad\;\;\; \\
{\cal M}_I^{12} & \sim & 
-i \int_{t_i}^{t_1} dt \, \delta_I(t) \, - \,
i \int_{t_i}^{t_1} dt \int_{t_i}^t dt'
\Bigl[ \, \gamma_I(t) \, \delta_I(t') - 
\delta_I(t) \, \gamma_I(t') \, \Bigr] 
\; , \\
{\cal M}_I^{21} & \sim &
-i \int_{t_i}^{t_1} dt \, \beta_I(t) \, - \,
i \int_{t_i}^{t_1} dt \int_{t_i}^t dt'
\Bigl[\, \beta_I(t) \, \gamma_I(t') - 
\gamma_I(t) \, \beta_I(t') \, \Bigr]
\; , \\
{\cal M}_I^{22} & \sim & 
1 \, - \, \int_{t_i}^{t_1} dt \, \gamma_I(t) \, + \,
\int_{t_i}^{t_1} dt \int_{t_i}^t dt'
\Bigl[ \, \gamma_I(t) \, \gamma_I(t') -
\beta_I(t) \, \delta_I(t') \, \Bigr] 
\; ,
\end{eqnarray}
where the coefficient functions are:
\begin{eqnarray}
\beta_I(t) & = &
\frac{\pi \dot{\nu_I} \; b_{\nu_I}(-\frac{x}{q_I})}
{2 \sin^2(\nu_I \pi)} 
\;\; , \\
\gamma_I(t) & = & 
\frac{\pi \dot{\nu_I} \; c_{\nu_I}(-\frac{x}{q_I})}
{4 \sin(\nu_I \pi)} 
\;\; , \\
\delta_I(t,k) & = &
\frac{\pi \dot{\nu}_I}{2} \;
d_{\nu_I}(\scriptstyle{-\frac{x}{q_I}}) 
\;\; .
\end{eqnarray}

\section{Appendix: Ultra Slow Roll Inflation}

Consider an inflaton potential like that depicted in Fig.~4 
and suppose inflation begins with the scalar to the right of 
the flat portion. Once the scalar rolls into the flat region 
its background equation of motion becomes:
\begin{equation}
\ddot{\varphi}_0 + 3 H \dot{\varphi}_0 \; = \; 0
\;\; .
\end{equation}
This can be integrated to give an exact expression for the 
scalar's time derivative in terms of its value at the 
beginning of the flat region:
\begin{equation}
\dot{\varphi}_0(t) \; = \;
\dot{\varphi}_f \, 
\Biggl( \frac{a_f}{a(t)} \Biggr)^3 
< \; 0 
\;\; .  \label{phiusr}
\end{equation}
If the scalar has enough kinetic energy it can roll through 
the flat region, and then on down its potential. The condition 
for this to happen is:
\begin{equation}
-\int_{t_f}^{\infty} dt \; \dot{\varphi}_f 
\Biggl( \frac{a_f}{a(t)} \Biggr)^3 \; \simeq \;
-\frac{\dot{\varphi}_f}{3 H_f} \; > \;
\varphi_f - \varphi_e 
\;\; .
\end{equation}
We shall assume this and study the scalar power spectrum for 
modes which experience first horizon crossing while the scalar 
is on the flat section.

\begin{figure}
\centerline{\epsfig{file=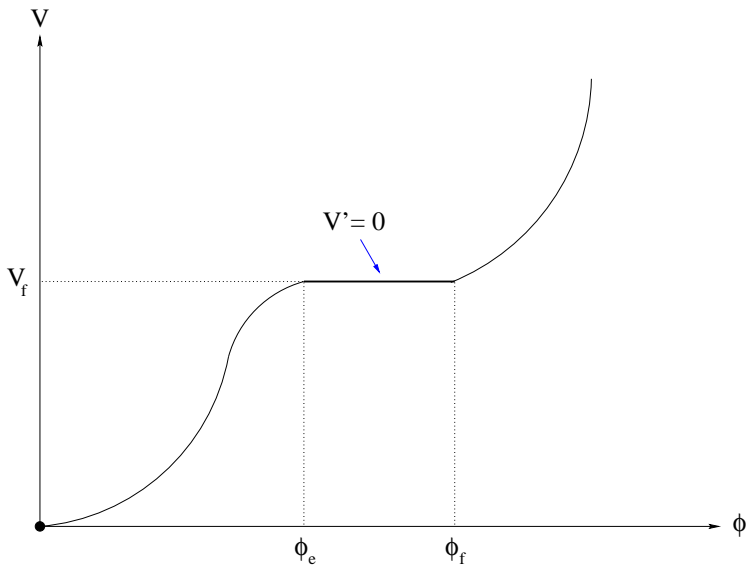,height=2.9in}}
\caption{\footnotesize The scalar potential associated with 
a phase of ultra slow roll inflation. 
\break \mbox{} \hspace{1.9cm} 
In the region $\varphi_e \leq \varphi \leq \varphi_f$ the 
potential is exactly flat with value $V_f$.}
\end{figure}

In the flat region all derivatives of the potential vanish, 
so all of the slow roll parameters are zero. Although the 
scalar is rolling ever more slowly -- hence the name -- 
this is a situation in which the conventional slow roll 
approximation completely breaks down. In fact the slow roll 
prediction (\ref{srpred}) for the scalar power spectrum 
actually diverges! The difficulty of reconciling this with 
a system which is approaching a pure de Sitter phase was 
the occasion of much reflection by Grishchuk 
\cite{grishchuk}. We shall see that ${\cal P}_{\rm SW}(k)$ 
is finite, but that it can become quite large.

By adding the background Einstein equations 
(\ref{E00}-\ref{Eij}) and then substituting (\ref{phiusr}) 
one finds:
\begin{equation}
1 + q(t) \; = \; 
4 \pi G \; \frac{\dot{\varphi}_0^2}{H^2} \; = \;
4 \pi G \; \Biggl( \frac{\dot{\varphi}_f}{H(t)} \Biggr)^2 
\, \Biggl( \frac{a_f}{a(t)} \Biggr)^6
\;\; . \label{q(t)}
\end{equation}
During inflation the deceleration parameter is typically 
near $-1$, but the fact that it approaches this value 
exponentially fast during the ultra slow roll phase makes 
a crucial change in the parameter $r(t)$ defined in 
(\ref{rdef}):
\begin{equation}
r(t) \; \equiv \; 
\frac1{H} \; \frac{d}{dt} \ln\Bigl(\sqrt{1 + q} \,\Bigr) 
\; = \; -3 - \frac{\dot{H}}{H^2} 
\; = \; -2 + q(t) 
\;\; . \label{r(t)}
\end{equation}
Although $r(t)$ is near zero for typical models of inflation, 
we see that it is nearly $-3$ during the ultra slow roll 
phase. It is simple enough to obtain an exact expression 
as well for its derivative during this phase:
\begin{equation}
\frac{\dot{r}}{H} \; = \;
\frac{\dot{q}}{H} \; = \;
2 (1 + q) \times 
\frac1{2H} \, \frac{\dot{q}}{1 + q} \; = \;
-2 (2 - q) (1 + q) 
\;\; .
\end{equation}
Note that this quantity is nearly zero, both for typical 
inflation and during ultra slow roll phase.

It is now straightforward to evaluate our factor 
${\cal C}_{1C}(k)$ that depends upon the system's state 
at horizon crossing. Substituting in (\ref{qC}) gives the 
following result during the ultra slow roll phase:
\begin{equation}
q_C(t) \; = - \frac{q}{q - 1} - \frac{2 (2 - q)(1 + q)}{(q-1)^2} \; = \;
1 + \frac1{q-1} - \frac4{(q-1)^2} \;\; .
\end{equation}
Although $q_C(t) \simeq +1$ in typical models of inflation, 
we see that it rapidly approaches $-\frac12$ during the 
ultra slow roll phase. Evaluating (\ref{C1C}) for 
$q_{1C} = -\frac12$ and $r_1 = -3$ gives:
\begin{equation}
{\cal C}_{1C}(k) \,
\Bigg\vert_{\, r_1 \, = \, -3}^{\, q_{1C} \, = \, -\frac12} 
\; = \; \frac1{4 \sqrt{\pi}} \; 
\Gamma \Bigl( -\frac32 \Bigr) \; = \; \frac13 \;\; .
\end{equation}

To estimate the evolution-dependent factor 
${\cal C}_{iC}(k)$ we make the reasonable assumption 
that the system goes suddenly from $q_C \simeq +1$ to 
$q_C \simeq -\frac12$. In this case the transfer matrix 
is determined by matching the mode functions and their 
first time derivatives at the onset of the flat region:
\footnote{In accordance with the definition (\ref{xdef}),
$x_f \equiv \frac{k}{H_f a_f}$.} 
\begin{eqnarray}
\Bigl( -i J_{-\frac52}(- x_f) \, , \, 
J_{\frac52}(- x_f) \Bigr) 
\left(\matrix{{\cal M}_C^{11} & {\cal M}_C^{12} \cr 
{\cal M}_C^{21} & {\cal M}_C^{22}} \right) \!\! & = & \!\! 
\Bigl( i J_{\frac12}(-x_f) \, , \,
J_{-\frac12}(-x_f) \Bigr) ,
\\
\Bigl( -i J'_{-\frac52}(- x_f) \, , \, 
J'_{\frac52}(- x_f) \Bigr) 
\left(\matrix{{\cal M}_C^{11} & {\cal M}_C^{12} \cr {\cal M}_C^{21} & 
{\cal M}_C^{22}} \right) \!\! & = & \!\! 
\Bigl( i J'_{\frac12}(-x_f) \, , \,
J'_{-\frac12}(-x_f) \Bigr) .
\;\;\;\;\;\;\;\;
\end{eqnarray}
The matrix elements needed for the scalar power spectrum 
are:
\begin{eqnarray}
{\cal M}_C^{21} & = & 
\frac{i \pi \, x_f}{2} \,
\Bigl[ \, -J_{-\frac52}(- x_f) \,\, J'_{\frac12}(-x_f) 
\; + \; J'_{-\frac52}(- x_f) \,\, J_{\frac12}(-x_f) \, \Bigr] 
\;\; , \\
{\cal M}_C^{22} & = & 
\frac{\pi \, x_f}{2} \,
\Bigl[ \, -J_{-\frac52}(- x_f) \,\, J'_{-\frac12}(-x_f) 
\; + \; J'_{-\frac52}(- x_f) \,\, J_{-\frac12}(-x_f) \, \Bigr ]
\;\; . \qquad
\end{eqnarray}
Substituting in (\ref{CiC}) with $q_{iC} = +1$ we obtain:
\begin{eqnarray}
{\cal C}_{iC}(k) & = & 
{\cal M}^{21}_C \, + \, {\cal M}^{22}_C 
\;\; , \\
& = & 
\sqrt{\frac{\pi x_f}2} \; e^{-i x_f} \, 
\Biggl\{  
- \Biggl[ \, 1 - \frac{i}{2 x_f} \, \Biggr] \, 
J_{-\frac52}(- x_f) \, - \, i \, J'_{-\frac52}(- x_f) 
\Biggr\} 
\;\; , \qquad \\
& = & 
-1 + \frac3{x_f} + \frac{3i}{x_f^2} - e^{-i x_f} \,
\Biggl\{ \frac3{x_f^2} \sin(x_f) + \frac6{x_f^3} \cos(x_f)
\Biggr\} \;\; . 
\end{eqnarray}

Because first horizon crossing occurs after the scalar 
has rolled onto the flat region we can assume $x_f > 1$. 
It is not safe to assume $x \gg 1$ because some modes 
will experience horizon crossing soon after the ultra 
slow roll phase begins. The power spectrum of these modes 
will deviate much more from scale invariance than is 
typically the case. Although the flat region must be 
narrow enough that the scalar can roll across, this 
process can be tuned to require an arbitrarily long time. 
For modes which experience horizon crossing long after 
the onset of the ultra slow roll phase, one can assume 
$x_f \gg 1$, in which case:
\begin{equation}
x_f \gg 1 
\quad \Longrightarrow\quad
\Vert \, {\cal C}_{iC}(k) \, \Vert^2 \; \simeq \; 1 \;\; . 
\end{equation}

We constructed this model as an exotic system in which 
the slow roll paradigm completely breaks down. However, 
it has two other properties worthy of note. The first is 
that, although our prediction (\ref{Pphi}) for the scalar 
power spectrum remains finite, it can become quite large 
owing to the inverse factor of $(1 + q_1)$. We have seen 
from (\ref{q(t)}) that $(1 + q(t))$ approaches zero 
exponentially fast. It seems inevitable that back-reaction 
must eventually become significant if the ultra slow roll 
phase is protracted.

The second interesting property of this model is that the 
anisotropies generated during the ultra slow roll phase 
are entirely due to scalar kinetic energy. The potential 
is completely flat so the only possible fluctuations derive 
from the gravitational response to kinetic energy. This is 
usually dismissed as negligible but we have just seen that 
it can drive an enormously strong effect as the system 
approaches de Sitter inflation. This suggests that one 
might expect a similarly strong effect from gravitons -- 
the combination of two of which can produce a scalar -- 
if the computation were carried to next order in the weak 
field expansion.

\end{document}